# Mechanical couplings of 3D lattice materials discovered by micropolar elasticity and geometric symmetry

Zhiming Cui, Zhihao Yuan, Jaehyung Ju

UM-SJTU Joint Institute, Shanghai Jiao Tong University, Shanghai 200240, China

**Abstract**: Similar to Poisson's effect, mechanical coupling is a directional indirect response by a directional input loading. With the advance in manufacturing techniques of 3D complex geometry, architected materials with unit cells of finite volume rather than a point yield more degrees of freedom and foster exotic mechanical couplings such as axial–shear, axial–rotation, axial–bending, and axial–twisting. However, most structural materials have been built by the ad hoc design of mechanical couplings without theoretical support of elasticity, which does not provide general guidelines for mechanical couplings. Moreover, no comprehensive study of all the mechanical couplings of 3D lattices with symmetry operations has been undertaken. Therefore, we construct the decoupled micropolar elasticity tensor of 3D lattices to identify individual mechanical couplings correlated with the point groups. The decoupled micropolar elasticity tensors, classified with 32 point groups, provide 15 mechanical couplings for 3D lattices. Our findings help provide solid theoretical guidelines for the mechanical couplings of 3D structural materials with potential applications in various areas, including active metamaterials, sensors, actuators, elastic waveguides, and acoustics.

**Keywords**: mechanical coupling, point groups, symmetry operation, micropolar elasticity, mechanical metamaterials

## I. Introduction

Since the advent of structural materials such as foams and lattice materials, numerous studies on axial–axial coupling, the so-called Poisson's effect, of structural materials have been conducted, e.g., zero and negative Poisson's ratios [1–6]. There are a few other couplings available in lattice materials: axial–rotation [7], axial–shear [8], axial–twisting [9,10], and axial–bending [11] couplings. Notably, most couplings are in 2D lattice materials. Only one mechanical coupling is available in 3D lattices, e.g., an axial–twisting [9,10]. Moreover, most couplings were secured based on ad hoc design without theoretical support of elasticity, which cannot provide a rational design guideline for mechanical coupling in lattice materials.

Mechanical coupling is a phenomenon whereby one input loading mode affects the other output deformation mode. Poisson's effect is the well-known mechanical coupling discovered two centuries ago that describes the interaction between a longitudinal input loading and the corresponding lateral output deformation [12]. In addition to Poisson's effect, so-called axial–axial coupling, the anisotropic Cauchy elasticity, describes other mechanical couplings such as axial–shear [13] and shear–shear [14] couplings in its constitutive equation [14]. In the Cauchy elasticity, a point-like infinitesimally small volume serves as the unit cell; it can displace but not bend and rotate, missing a certain degree of freedom in artificially designed lattices where the unit cells have a finite volume rather than a point. Even though it may not be strictly called a material's constitutive model, the classical lamination theory (**CLT**) [15,16] can also describe a certain level of mechanical couplings such as axial–shear, axial–bending, shear–twisting,

bending–bending, and bending–twisting couplings. **However**, similar to the Cauchy elasticity, CLT still lacks full coupling modes in 3D space due to the lack of complex micro deformation of a lamina.

Symmetry is a unique feature of lattice geometry distinguished from the continuum and can serve as a critical tool for designing various mechanical couplings. In 2D planar lattices, chirality refers to lattices with broken mirror symmetries [17,18], leading to axial–rotation [7] and axial–shear [8] couplings. A 2D lattice that does not superimpose after rotating by 180° is called non-centrosymmetric [17], leading to an axial–bending coupling [11]. In 3D materials, Lakes and Benedict developed a theoretical layout of an axial–twisting coupling of non-centrosymmetric isotropic materials in 1982 [10], which was demonstrated with experiments in 2017 [9]. However, such classification as chiral and non-centrosymmetric is rather too broad of a concept and cannot precisely describe all coupling effects in lattice materials. Notably, there are four symmetry operations in 3D lattices — translation, rotation, mirror, and inversion symmetries [19]. The chirality is characterized by only the rotational symmetry: two-fold, three-fold, four-fold, and six-fold rotational symmetries. Non-centrosymmetry is equivalent to inversional symmetry, which is independent of chirality. Moreover, mirror symmetry is not considered in such classification.

Micropolar elasticity has been adapted to describe a finite unit cell's complex motion with micro-rotation [20,21]. This theory can potentially cover more complex deformation patterns in 2D and 3D lattice materials while covering both the Cauchy and couple strains. The Cauchy strain includes the axial and shear strains, where the shear strain can be decomposed into pure shear and joint rotation [22]. The couple strain includes bending and twisting [20,21]. Therefore, there are five independent deformation patterns in 3D micropolar elasticity: axial deformation, pure shear, joint rotation, bending, and twisting. They can potentially produce 15 mechanical couplings with a combination of the independent deformations, which will be covered in this work. Similar to the previous work on 2D lattices [23], we construct a modified micropolar elasticity tensor by decoupling pure shear and nodal rotation in the Cauchy shear strain, revealing the mechanical coupling effects in the elasticity tensor. Following Neumann's principle [19], we reduce the elasticity tensor for varying symmetry operations and the point group theory in crystallography, producing possible coupling effects of 3D lattices for 32 point groups. We verify our theory by designing several 3D lattices and demonstrating couplings through finite element (FE) simulations.

## II.    Decoupling of micropolar elasticity tensor

In the 3D micropolar elasticity, every element has six degrees of freedom, including three translations $u_i$ and three rotations $\phi_i$, $i = 1, 2, 3$. The constitutive equations of the micropolar elasticity are given by

$$\sigma_{ij} = C_{ijkl}\varepsilon_{kl} + B_{ijkl}\kappa_{kl}$$
$$m_{ij} = B_{klij}\varepsilon_{kl} + D_{ijkl}\kappa_{kl}, \quad (1)$$

where $\sigma_{ij}$ is the Cauchy stress tensor, $m_{ij}$ is the couple stress tensor, $\varepsilon_{ij}$ is the Cauchy strain tensor, and $\kappa_{ij}$ is the couple strain tensor.

The tensor form can be written as the matrix form with the micropolar elasticity tensor **Q** of $18 \times 18$, as shown in Equation (2). Notably, the micropolar elasticity tensor **Q** can be decomposed into the **C**, **B**, and **D** sub-matrices.

$$\boldsymbol{\sigma} = \mathbf{Q} \cdot \boldsymbol{\varepsilon}$$

$$\begin{Bmatrix}\sigma_{11}\\\sigma_{22}\\\sigma_{33}\\\sigma_{23}\\\sigma_{13}\\\sigma_{12}\\\sigma_{32}\\\sigma_{31}\\\sigma_{21}\\m_{11}\\m_{22}\\m_{33}\\m_{23}\\m_{13}\\m_{12}\\m_{32}\\m_{31}\\m_{21}\end{Bmatrix} = \begin{bmatrix} C_{11} & C_{12} & C_{13} & C_{14} & C_{15} & C_{16} & C_{17} & C_{18} & C_{19} & B_{11} & B_{12} & B_{13} & B_{14} & B_{15} & B_{16} & B_{17} & B_{18} & B_{19}\\ & C_{22} & C_{23} & C_{24} & C_{25} & C_{26} & C_{27} & C_{28} & C_{29} & B_{21} & B_{22} & B_{23} & B_{24} & B_{25} & B_{26} & B_{27} & B_{28} & B_{29}\\ & & C_{33} & C_{34} & C_{35} & C_{36} & C_{37} & C_{38} & C_{39} & B_{31} & B_{32} & B_{33} & B_{34} & B_{35} & B_{36} & B_{37} & B_{38} & B_{39}\\ & & & C_{44} & C_{45} & C_{46} & C_{47} & C_{48} & C_{49} & B_{41} & B_{42} & B_{43} & B_{44} & B_{45} & B_{46} & B_{47} & B_{48} & B_{49}\\ & & & & C_{55} & C_{56} & C_{57} & C_{58} & C_{59} & B_{51} & B_{52} & B_{53} & B_{54} & B_{55} & B_{56} & B_{57} & B_{58} & B_{59}\\ & & & & & C_{66} & C_{67} & C_{68} & C_{69} & B_{61} & B_{62} & B_{63} & B_{64} & B_{65} & B_{66} & B_{67} & B_{68} & B_{69}\\ & & & & & & C_{77} & C_{78} & C_{79} & B_{71} & B_{72} & B_{73} & B_{74} & B_{75} & B_{76} & B_{77} & B_{78} & B_{79}\\ & & & & & & & C_{88} & C_{89} & B_{81} & B_{82} & B_{83} & B_{84} & B_{85} & B_{86} & B_{87} & B_{88} & B_{89}\\ & & & & & & & & C_{99} & B_{91} & B_{92} & B_{93} & B_{94} & B_{95} & B_{96} & B_{97} & B_{98} & B_{99}\\ & & & & & & & & & D_{11} & D_{12} & D_{13} & D_{14} & D_{15} & D_{16} & D_{17} & D_{18} & D_{19}\\ & & & & & & & & & & D_{22} & D_{23} & D_{24} & D_{25} & D_{26} & D_{27} & D_{28} & D_{29}\\ & & & & & & & & & & & D_{33} & D_{34} & D_{35} & D_{36} & D_{37} & D_{38} & D_{39}\\ & & & \text{Symmetric} & & & & & & & & & D_{44} & D_{45} & D_{46} & D_{47} & D_{48} & D_{49}\\ & & & & & & & & & & & & & D_{55} & D_{56} & D_{57} & D_{58} & D_{59}\\ & & & & & & & & & & & & & & D_{66} & D_{67} & D_{68} & D_{69}\\ & & & & & & & & & & & & & & & D_{77} & D_{78} & D_{79}\\ & & & & & & & & & & & & & & & & D_{88} & D_{89}\\ & & & & & & & & & & & & & & & & & D_{99}\end{bmatrix} \begin{Bmatrix}\varepsilon_{11}\\\varepsilon_{22}\\\varepsilon_{33}\\\varepsilon_{23}\\\varepsilon_{13}\\\varepsilon_{12}\\\varepsilon_{32}\\\varepsilon_{31}\\\varepsilon_{21}\\\kappa_{11}\\\kappa_{22}\\\kappa_{33}\\\kappa_{23}\\\kappa_{13}\\\kappa_{12}\\\kappa_{32}\\\kappa_{31}\\\kappa_{21}\end{Bmatrix}$$

(2)

Note that we assemble $\sigma_{ij}$ with $m_{ij}$ for a generalized stress vector form in Equation (3); $\boldsymbol{\sigma} = \{\sigma_{11}, \sigma_{22}, \sigma_{33}, \sigma_{23}, \sigma_{13}, \sigma_{12}, \sigma_{32}, \sigma_{31}, \sigma_{21}, m_{11}, m_{22}, m_{33}, m_{23}, m_{13}, m_{12}, m_{32}, m_{31}, m_{21}\}^T = \{\sigma_m\}^T$, where $m = 1, \ldots, 18$. Similarly, we assemble $\varepsilon_{ij}$ with $\kappa_{ij}$ for a generalized strain vector form in Equation (3); $\boldsymbol{\varepsilon} = \{\varepsilon_{11}, \varepsilon_{22}, \varepsilon_{33}, \varepsilon_{23}, \varepsilon_{13}, \varepsilon_{12}, \varepsilon_{32}, \varepsilon_{31}, \varepsilon_{21}, \kappa_{11}, \kappa_{22}, \kappa_{33}, \kappa_{23}, \kappa_{13}, \kappa_{12}, \kappa_{32}, \kappa_{31}, \kappa_{21}\}^T = \{\varepsilon_m\}^T$, where $m = 1, \ldots, 18$.

In the micropolar elasticity, one can specify strain nontraditionally: the Cauchy strain for axial and shear modes and the couple strain for bending and twisting modes. The Cauchy and couple strains are defined as, respectively,

$$\varepsilon_{ij} = u_{j,i} - e_{ijk}\phi_k$$
$$\kappa_{ij} = \phi_{j,i}, \tag{3}$$

where $e_{ijk}$ is the permutation tensor, and the comma in the subscript denotes a partial differentiation for the subsequent spatial variable. For $i = j$, the Cauchy strain $\varepsilon_{ij}$ represents normal strain with $e_{ijk} = 0$. $\varepsilon_{ij}$ implies a combination of shear strain and rotation for $i \neq j$ and $e_{ijk} \neq 0$. Similarly, the couple strain $\kappa_{ij}$ represents twisting for when $i = j$ [9,24]. However, $\kappa_{ij}$ implies a bending deformation for $i \neq j$.

Note also that $\sigma_{ij}$ and $\varepsilon_{ij}$ in Equations (1) and (3) are asymmetric tensors. To decouple $\varepsilon_{ij}$ into the shear and rotation modes, we decompose the asymmetric stress (or stress) tensor into symmetric and anti-symmetric parts [22,25],

$$\sigma_{ij} = S_{ij} + T_{ji}, \quad \varepsilon_{ij} = E_{ij} + A_{ji}, \tag{4}$$

where

$$S_{ij} = \frac{\sigma_{ij} + \sigma_{ji}}{2}, \quad T_{ij} = \frac{\sigma_{ij} - \sigma_{ji}}{2}$$
$$E_{ij} = \frac{\varepsilon_{ij} + \varepsilon_{ji}}{2}, \quad A_{ij} = \frac{\varepsilon_{ij} - \varepsilon_{ji}}{2}. \tag{5}$$

Apparently, $S_{ij} = S_{ji}$ and $E_{ij} = E_{ji}$ are symmetric. $T_{ij} = -T_{ji}$ and $A_{ij} = -A_{ji}$ are anti-symmetric.

Note that

$$E_{ij} = E_{ji} = \frac{\varepsilon_{ij} + \varepsilon_{ji}}{2} = \frac{u_{j,i} + u_{i,j}}{2} = \frac{\gamma_{ij}}{2}$$

$$A_{ij} = -A_{ji} = \frac{\varepsilon_{ij} - \varepsilon_{ji}}{2} = \frac{u_{j,i} + u_{i,j}}{2} - e_{ijk}\phi_k = e_{ijk}(\Omega_k - \phi_k), \tag{6}$$

where $\Omega_k = \frac{1}{2}e_{ijk}u_{j,i}$ is the macro-rotation produced by the displacement field. The physical meanings of $E_{ij}$ and $A_{ij}$ are relatively straightforward with Equation (6), i.e., $E_{ij}$ is half the engineering shear strain $\gamma_{ij}$, and $A_{ij}$ represents the local rotation at one point, as illustrated in Figure 1.

The macro-rotation $\Omega_k$ differs from the local rotation $A_{ij}$. For example, consider the deformation of a beam strut in a lattice, as shown in Figure 1. $\Omega_k$ is explicitly dependent on the joint's displacement, whereas the local rotation $A_{ij}$ is affected by the beam's shape and the moments applied. The nodal rotation $\phi_k$ is a combination of $\Omega_k$ and $A_{ji}$, representing a total rotation of the joint.

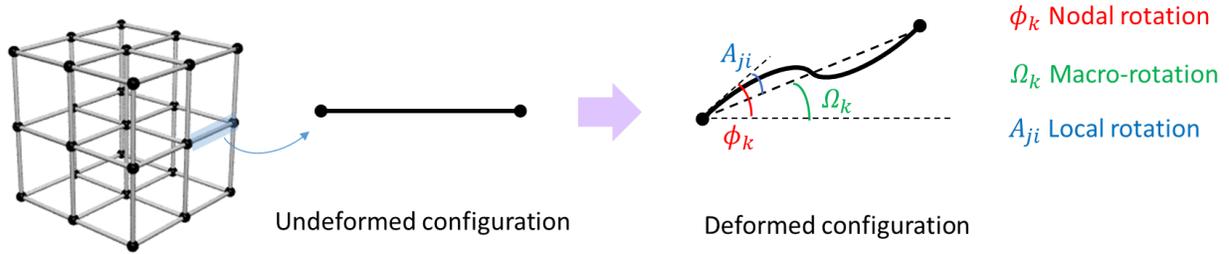

Figure 1. Physical meanings of nodal rotation $\phi_k (= \Omega_k - e_{ijk}A_{ij})$, macro-rotation $\Omega_k$, and local rotation $A_{ji}$.

We can build a constitutive relation of 3D lattices with the decoupled Cauchy stress and strain tensors in Equations (4):

$$\mathbf{S} = \mathbf{q} \cdot \mathbf{E}$$

$$\begin{Bmatrix} S_{11} \\ S_{22} \\ S_{33} \\ S_{23} \\ S_{13} \\ S_{12} \\ T_{23} \\ T_{13} \\ T_{12} \\ m_{11} \\ m_{22} \\ m_{33} \\ m_{23} \\ m_{13} \\ m_{12} \\ m_{32} \\ m_{31} \\ m_{21} \end{Bmatrix} = \begin{bmatrix} c_{11} & c_{12} & c_{13} & c_{14} & c_{15} & c_{16} & c_{17} & c_{18} & c_{19} & b_{11} & b_{12} & b_{13} & b_{14} & b_{15} & b_{16} & b_{17} & b_{18} & b_{19} \\ & c_{22} & c_{23} & c_{24} & c_{25} & c_{26} & c_{27} & c_{28} & c_{29} & b_{21} & b_{22} & b_{23} & b_{24} & b_{25} & b_{26} & b_{27} & b_{28} & b_{29} \\ & & c_{33} & c_{34} & c_{35} & c_{36} & c_{37} & c_{38} & c_{39} & b_{31} & b_{32} & b_{33} & b_{34} & b_{35} & b_{36} & b_{37} & b_{38} & b_{39} \\ & & & c_{44} & c_{45} & c_{46} & c_{47} & c_{48} & c_{49} & b_{41} & b_{42} & b_{43} & b_{44} & b_{45} & b_{46} & b_{47} & b_{48} & b_{49} \\ & & & & c_{55} & c_{56} & c_{57} & c_{58} & c_{59} & b_{51} & b_{52} & b_{53} & b_{54} & b_{55} & b_{56} & b_{57} & b_{58} & b_{59} \\ & & & & & c_{66} & c_{67} & c_{68} & c_{69} & b_{61} & b_{62} & b_{63} & b_{64} & b_{65} & b_{66} & b_{67} & b_{68} & b_{69} \\ & & & & & & c_{77} & c_{78} & c_{79} & b_{71} & b_{72} & b_{73} & b_{74} & b_{75} & b_{76} & b_{77} & b_{78} & b_{79} \\ & & & & & & & c_{88} & c_{89} & b_{81} & b_{82} & b_{83} & b_{84} & b_{85} & b_{86} & b_{87} & b_{88} & b_{89} \\ & & & & & & & & c_{99} & b_{91} & b_{92} & b_{93} & b_{94} & b_{95} & b_{96} & b_{97} & b_{98} & b_{99} \\ & & & & & & & & & d_{11} & d_{12} & d_{13} & d_{14} & d_{15} & d_{16} & d_{17} & d_{18} & d_{19} \\ & & & & & & & & & & d_{22} & d_{23} & d_{24} & d_{25} & d_{26} & d_{27} & d_{28} & d_{29} \\ & & & \text{Symmetric} & & & & & & & & d_{33} & d_{34} & d_{35} & d_{36} & d_{37} & d_{38} & d_{39} \\ & & & & & & & & & & & & d_{44} & d_{45} & d_{46} & d_{47} & d_{48} & d_{49} \\ & & & & & & & & & & & & & d_{55} & d_{56} & d_{57} & d_{58} & d_{59} \\ & & & & & & & & & & & & & & d_{66} & d_{67} & d_{68} & d_{69} \\ & & & & & & & & & & & & & & & d_{77} & d_{78} & d_{79} \\ & & & & & & & & & & & & & & & & d_{88} & d_{89} \\ & & & & & & & & & & & & & & & & & d_{99} \end{bmatrix} \begin{Bmatrix} E_{11} \\ E_{22} \\ E_{33} \\ 2E_{23} \\ 2E_{13} \\ 2E_{12} \\ 2A_{23} \\ 2A_{13} \\ 2A_{12} \\ \kappa_{11} \\ \kappa_{22} \\ \kappa_{33} \\ \kappa_{23} \\ \kappa_{13} \\ \kappa_{12} \\ \kappa_{32} \\ \kappa_{31} \\ \kappa_{21} \end{Bmatrix}$$

(7)

An 18 × 18 matrix form of the decoupled elasticity tensor **q** in Equation (7) can be decomposed with the **c**, **b**, and **d** sub-matrices, which is comparable to the coupled elasticity tensor **Q** of Equation (2). Using Equations (4) and (5), one can obtain the correlation between the two elasticity tensors **Q** to **q**. Because of the page limit of the main text, we provide a MATHEMATICA code to transform **Q** to **q** in [26]. Note that we assemble $S_{ij}$, $T_{ij}$, and $m_{ij}$ for a generalized stress vector form in Equation (7); $\mathbf{S} = \{S_{11}, S_{22}, S_{33}, S_{23}, S_{13}, S_{12}, T_{23}, T_{13}, T_{12}, m_{11}, m_{22}, m_{33}, m_{23}, m_{13}, m_{12}, m_{32}, m_{31}, m_{21}\}^T = \{S_m\}^T$, where $m = 1, \ldots, 18$. Similarly, we assemble $E_{ij}$, $A_{ij}$, and $\kappa_{ij}$ for a generalized strain vector form in Equation (7); $\boldsymbol{\varepsilon} = \{E_{11}, E_{22}, E_{33}, 2E_{23}, 2E_{13}, 2E_{12}, 2A_{23}, 2A_{13}, 2A_{12}, \kappa_{11}, \kappa_{22}, \kappa_{33}, \kappa_{23}, \kappa_{13}, \kappa_{12}, \kappa_{32}, \kappa_{31}, \kappa_{21}\}^T = \{E_m\}^T$, where $m = 1, \ldots, 18$. Note that we use $2E_{23}, 2E_{13}, 2E_{12}, 2A_{23}, 2A_{13}$, and $2A_{12}$ to make **q** symmetric. In addition, note that the first six rows and columns of **q** match the classical Cauchy elasticity tensor.

The highlight of indirect strain components of Equation (7) in Figure 2 helps us identify mechanical couplings in **q**. The diagonal components of **q** are related to a direct resistance to applied input strains; e.g., Young's modulus ($c_{11}$, $c_{22}$, and $c_{33}$), shear modulus ($c_{44}$, $c_{55}$, and $c_{66}$), and bending stiffness ($d_{11}$, $d_{22}$, and $d_{33}$). However, one can identify the coupling effects from the non-diagonal terms of **q**, as highlighted in Figure 2. There are **five** independent deformation modes — axial, shear, rotation, twisting, and bending, which lead to **15** mechanical coupling effects: axial–axial (**AA**), axial–shear (**AS**), axial–rotation (**AR**), axial–twisting (**AT**), axial–bending (**AB**), shear–shear (**SS**), shear–rotation (**SR**), shear–twisting (**ST**), shear–bending (**SB**), rotation–rotation (**RR**), rotation–twisting (**RT**), rotation–bending (**RB**), twisting–twisting (**TT**), twisting–bending (**TB**), and bending–bending (**BB**) couplings. The AA coupling is called Poisson's ratio [12]. The AS, AR, AB, AT, and SS couplings were implemented with lattice design [7–9,11,13,14,24]; however, the remaining nine couplings — SR, ST, SB, RR, RT, RB, TT, TB, and BB couplings have not yet been implemented for structural applications.

The classical Cauchy elasticity can describe AA, AS, and SS couplings. CLT can also provide AS, AB, AT, SB, ST, BB, and TB couplings [16]. However, this work's decomposition of the micropolar elasticity tensor can provide new couplings; one can identify AR, SR, RR, RT, RB, TT, and BB couplings uniquely by the micropolar elasticity. The number of components to affect the individual couplings are different between CLT and this work; e.g., AB coupling is related to four components for CLT [16] but 18 components in this work. Note that each component in a coupling has independent modes; an AA coupling has three components, $c_{12}$, $c_{13}$, and $c_{23}$, which affect Poisson's effect $v_{12}$, $v_{13}$, and $v_{23}$, respectively. AA, SS, RR, and TT couplings have three components. AS, AR, AT, SR, ST, and RT couplings possess nine components. AB, SB, RB, and TB couplings have 18 components. Interestingly, a BB coupling has 15 components.

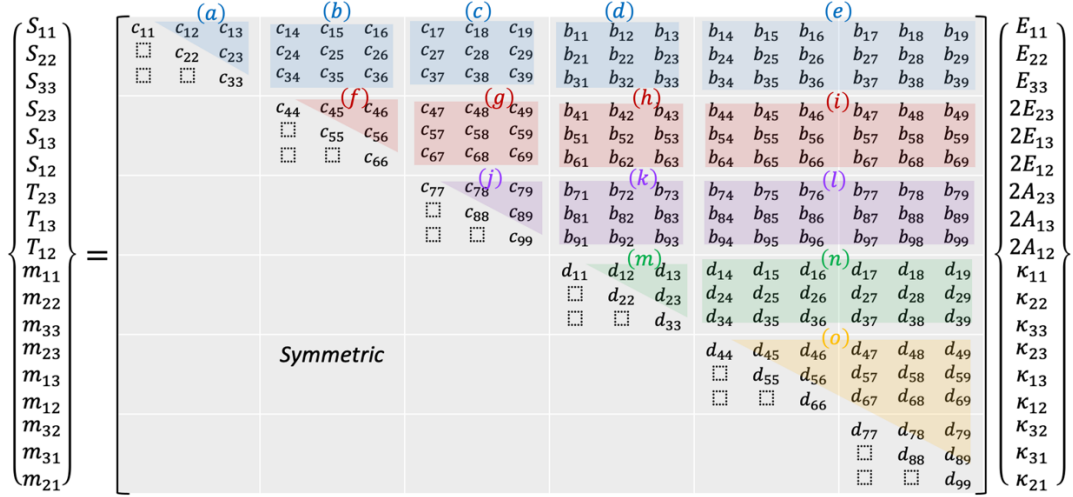

(a) Axial-Axial (AA)
(b) Axial-Shear (AS)       (f) Shear-Shear (SS)
(c) Axial-Rotation (AR)    (g) Shear-Rotation (SR)    (j) Rotation-Rotation (RR)
(d) Axial-Twist (AT)       (h) Shear-Twist (ST)       (k) Rotation-Twisting (RT)   (m) Twisting-Twisting (TT)
(e) Axial-Bending (AB)     (i) Shear-Bending (SB)     (l) Rotation-Bending (RB)    (n) Twisting-Bending (TB)    (o) Bending-Bending (BB)

Figure 2. Mechanical couplings in the decoupled micropolar elasticity tensor $\mathbf{q}$.

Figure 3 shows the 15 mechanical couplings of 3D lattices. Although each coupling has multiple configurations depending on the input loading directions, we only show one case for illustration. The AS coupling in Figure 3b has been found in 2D chiral square lattices [8,13,27]. The AR coupling in Figure 3c was found in a 2D chiral triangular lattice [7]. Recently, an AB coupling in Figure 3e was realized in a 2D non-centrosymmetric square lattice [11]. The AT coupling in Figure 3d has received attention as a unique feature in 3D lattices [9].

The BB coupling can be further divided into in-plane and out-of-plane modes. The in-plane mode refers to the coupling between $\kappa_{ik}$ and $\kappa_{jk}$, where the bending deformations are in the same plane. For example, the BB coupling in 2D lattices is an in-plane mode [11], as the involved bending curvatures are $\kappa_{13}$ and $\kappa_{23}$. The other BB couplings are in the out-of-plane mode, such as the BB coupling shown in Figure 3o. In terms of the components of $\mathbf{q}$, $d_{45}$, $d_{67}$, and $d_{89}$ represent an in-plane BB coupling, whereas the other components represent an out-of-plane BB coupling. Similarly, the SB coupling has both in-plane and out-of-plane modes.

Although Figure 3 illustrates independent coupling modes, multimodal coupling effects can appear in one lattice; e.g., an axial input loading can produce shear and twisting deformations simultaneously.

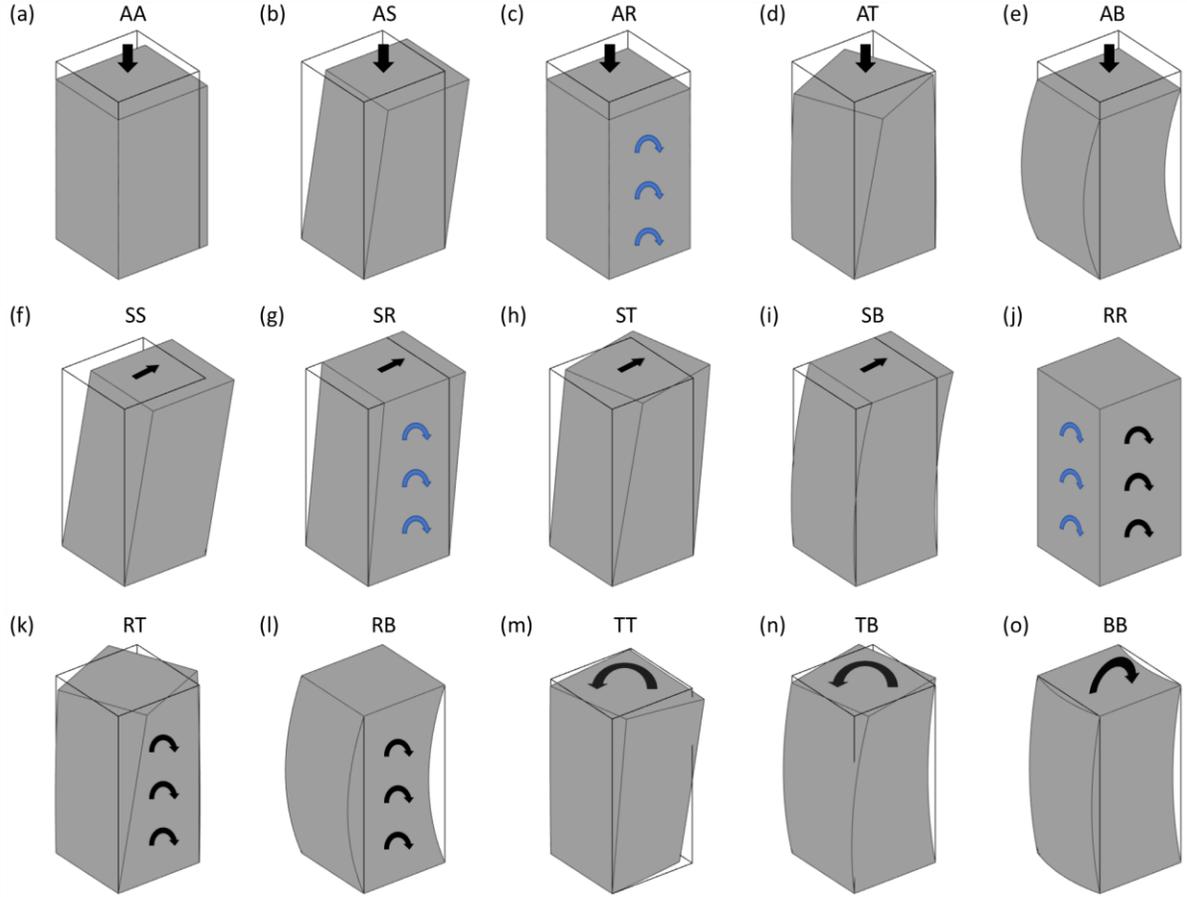

*Figure 3. Visual interpretation of 3D mechanical couplings: (a) axial–axial (AA), (b) axial–shear (AS), (c) axial–rotation (AR), (d) axial–twisting (AT), (e) axial–bending (AB), (f) shear–shear (SS), (g) shear–rotation (SR), (h) shear–twisting (ST), (i) shear–bending (SB), (j) rotation–rotation (RR), (k) rotation–twisting (RT), (l) rotation–bending (RB), (m) twisting–twisting (TT), (n) twisting–bending (TB), (o) bending–bending (BB) couplings.*

## III. Symmetry operations of 3D lattices

As shown in Figure 2, the fully anisotropic micropolar elasticity tensor has 171 independent components. A lattice with a certain degree of symmetry produces some zero off-diagonal components. There are 11 symmetry operations in 3D space, as summarized in Table 1. We use Neumann's principle to identify the nonzero terms of the micropolar elasticity tensor **Q** to analyze mechanical couplings [19]. Every symmetry operation corresponds to a coordinate transformation by the direction cosine $l_{ij}(=\mathbf{L})$ between unbarred ($\hat{\mathbf{e}}_j$) and barred ($\hat{\bar{\mathbf{e}}}_i$) coordinate bases:

$$\hat{\bar{\mathbf{e}}}_i = l_{ij}\hat{\mathbf{e}}_j. \tag{8}$$

Table 1. Transformation matrices and symbols for selected symmetry operations [19]; $Z_1$, $Z_2$, and $Z_3$ refer to the three principal axes.

| Symmetry operation | Symbol | Transformation matrix, $\mathbf{L}$ |
|---|---|---|
| Identity operator | 1 | $\begin{pmatrix} 1 & 0 & 0 \\ 0 & 1 & 0 \\ 0 & 0 & 1 \end{pmatrix}$ |
| Inversion center | $\bar{1}$ | $\begin{pmatrix} -1 & 0 & 0 \\ 0 & -1 & 0 \\ 0 & 0 & -1 \end{pmatrix}$ |
| Two-fold rotation parallel to $Z_1$ | $2 \parallel Z_1$ | $\begin{pmatrix} 1 & 0 & 0 \\ 0 & -1 & 0 \\ 0 & 0 & -1 \end{pmatrix}$ |
| Two-fold rotation parallel to $Z_2$ | $2 \parallel Z_2$ | $\begin{pmatrix} -1 & 0 & 0 \\ 0 & 1 & 0 \\ 0 & 0 & -1 \end{pmatrix}$ |
| Mirror perpendicular to $Z_1$ | $m \perp Z_1$ | $\begin{pmatrix} -1 & 0 & 0 \\ 0 & 1 & 0 \\ 0 & 0 & 1 \end{pmatrix}$ |
| Mirror perpendicular to $Z_2$ | $m \perp Z_2$ | $\begin{pmatrix} 1 & 0 & 0 \\ 0 & -1 & 0 \\ 0 & 0 & 1 \end{pmatrix}$ |
| Mirror perpendicular to $Z_3$ | $m \perp Z_3$ | $\begin{pmatrix} 1 & 0 & 0 \\ 0 & 1 & 0 \\ 0 & 0 & -1 \end{pmatrix}$ |
| Three-fold rotation parallel to $Z_3$ | $3 \parallel Z_3$ | $\begin{pmatrix} -1/2 & \sqrt{3}/2 & 0 \\ -\sqrt{3}/2 & -1/2 & 0 \\ 0 & 0 & 1 \end{pmatrix}$ |
| Three-fold rotation parallel to [1 1 1] | $3 \parallel [1\,1\,1]$ | $\begin{pmatrix} 0 & 1 & 0 \\ 0 & 0 & 1 \\ 1 & 0 & 0 \end{pmatrix}$ |
| Three-fold inversion axis parallel to $Z_3$ | $\bar{3} \parallel Z_3$ | $\begin{pmatrix} 1/2 & -\sqrt{3}/2 & 0 \\ \sqrt{3}/2 & 1/2 & 0 \\ 0 & 0 & -1 \end{pmatrix}$ |
| Three-fold inversion axis parallel to [1 1 1] in cubic crystals | $\bar{3} \parallel [1\,1\,1]$ | $\begin{pmatrix} 0 & -1 & 0 \\ 0 & 0 & -1 \\ -1 & 0 & 0 \end{pmatrix}$ |

According to the transformation law of second-order tensors [19], the stress and couple stress in Equation (1) transforms as

$$\bar{\sigma}_{ij} = l_{ik} l_{jl} \sigma_{kl}$$
$$\bar{m}_{ij} = |\mathbf{L}| l_{ik} l_{jl} m_{kl}, \qquad (9)$$

where $|\mathbf{L}|$ is the determinant of $\mathbf{L}$. Notably, $|\mathbf{L}| = -1$ for the symmetry operations involving mirror planes or inversion centers and $|\mathbf{L}| = 1$ for rotational symmetry. Note that the transformation principle of $\bar{\sigma}_{ij}$ and $\bar{m}_{ij}$ in Equation (9) are different because the Cauchy stress is a polar tensor while the couple stress is an axial tensor [28].

The transformation matrix $\mathbf{R}$ whose size is $18 \times 18$ can be defined from Equation (9):

$$\mathbf{R} = \begin{bmatrix} l_{ik}l_{jl} & 0 \\ 0 & |\mathbf{L}|l_{ik}l_{jl} \end{bmatrix}. \tag{10}$$

The full expression of $\mathbf{R}$ is shown in Section A1 of the Appendix.

The transformed stress in a vector form is given by

$$\begin{Bmatrix} \bar{\sigma}_{ij} \\ \bar{m}_{ij} \end{Bmatrix} = \mathbf{R} \cdot \begin{Bmatrix} \sigma_{ij} \\ m_{ij} \end{Bmatrix}. \tag{11}$$

The transformed strain is given by

$$\begin{Bmatrix} \bar{\varepsilon}_{kl} \\ \bar{\kappa}_{kl} \end{Bmatrix} = \mathbf{R}^{-T} \cdot \begin{Bmatrix} \varepsilon_{kl} \\ \kappa_{kl} \end{Bmatrix}. \tag{12}$$

Notably, $\mathbf{R}$ is an orthogonal transformation: $\mathbf{R}^{-T} = \mathbf{R}$.

The constitutive equation in the transformed coordinate is

$$\bar{\boldsymbol{\sigma}} = \bar{\mathbf{Q}} \cdot \bar{\boldsymbol{\varepsilon}}. \tag{13}$$

Substituting Equations (11) and (12) into Equation (13) gives:

$$\mathbf{R} \cdot \boldsymbol{\sigma} = \bar{\mathbf{Q}} \cdot \mathbf{R}^{-T} \cdot \boldsymbol{\varepsilon}, \tag{14}$$

i.e.,

$$\boldsymbol{\sigma} = \mathbf{R}^{-1} \cdot \bar{\mathbf{Q}} \cdot \mathbf{R}^{-T} \cdot \boldsymbol{\varepsilon}. \tag{15}$$

Note that the elasticity tensors in the original and transformed coordinates are given by

$$\mathbf{Q} = \mathbf{R}^{-1} \cdot \bar{\mathbf{Q}} \cdot \mathbf{R}^{-T} \tag{16}$$

$$\bar{\mathbf{Q}} = \mathbf{R} \cdot \mathbf{Q} \cdot \mathbf{R}^T. \tag{17}$$

According to the principle of symmetry [19], if the geometry is invariant under such transformation, the elasticity tensor should also be invariant, i.e.,

$$\bar{\mathbf{Q}} = \mathbf{Q}. \tag{18}$$

Solving Equation (18), by comparing the elasticity tensor before and after symmetry operation, one can identify the nonzero components of $\mathbf{Q}$. One can obtain the decoupled elasticity tensor $\mathbf{q}$ from $\mathbf{Q}$ via Equations (4) and (5). We show the detailed relation of components between in Section A2 of the Appendix.

## IV. Mechanical coupling in the point groups

In general, lattices contain more than one symmetry operation [17,19]. According to the symmetry principle, one can classify 3D lattices into the 32 point groups listed in Table 2, shown together with the corresponding symmetry elements, chirality, and centrosymmetry [17,19]. Note that chirality and non-centrosymmetry have been widely used to describe physical properties with a broken symmetry of lattices' geometry [9,29,30]. However, the use of chirality and centrosymmetry alone cannot fully describe the geometric symmetry of 3D lattices, as shown in Table 2. Therefore, the point group theory must be involved for comprehensive analysis of the symmetry of 3D lattices. Note that the chirality was misused in previous work [9], where the 3D lattice belongs to point group 432 without chirality.

Each point group has corresponding symmetry operations where we can find the mechanical coupling effects with Equation (16), followed by the decoupling process with Equation (5) to obtain **q**. Table 2 summarizes the potential coupling effects of transformed **q** by each symmetry operation.

It is apparent that the number of potential coupling effects decreases as the degree of symmetry increases; point group 1 with no symmetry has all 15 possible coupling effects. However, point group $m\bar{3}m$ with increased symmetry has AA, TT, and BB couplings. These three couplings, AA, TT, and BB, exist in every point group, even in the isotropic micropolar continuum in Table 8. The Cauchy elasticity can only describe AA, AS, and SS couplings [8,13]. Note also that the AA coupling exists in all point groups and isotropic materials [19], explaining the common observation of Poisson's effect among 15 coupling effects.

*Table 2. Potential coupling effects for the 32 point groups*

| Point group | Symmetry operation | Chiral | Centrosymmetric | Potential coupling effects |
|---|---|---|---|---|
| 1 | 1 | Yes | No | AA, AS, AR, AT, AB, SS, SR, ST, SB, RR, RT, RB, TT, TB, BB |
| $\bar{1}$ | $\bar{1}$ | No | Yes | AA, AS, AR, SS, SR, RR, TT, TB, BB |
| 2 | $2 \parallel Z_2$ | Yes | No | AA, AS, AR, AT, AB, SS, SR, ST, SB, RR, RT, RB, TT, TB, BB |
| $m$ | $m \perp Z_2$ | No | No | AA, AS, AR, AB, SS, SR, ST, SB, RR, RT, RB, TT, TB, BB |
| $2/m$ | $2 \parallel Z_2, m \perp Z_2$ | No | Yes | AA, AS, AR, SS, SR, RR, TT, TB, BB |
| 222 | $2 \parallel Z_1, 2 \parallel Z_2$ | Yes | No | AA, AT, SR, SB, RB, TT, BB |
| $mm2$ | $m \perp Z_1, m \perp Z_2$ | No | No | AA, AB, SR, ST, SB, RT, RB, TT, BB |
| $mmm$ | $m \perp Z_1, m \perp Z_2, m \perp Z_3$ | No | Yes | AA, SR, TT, BB |
| $4$ | $4 \parallel Z_3$ | Yes | No | AA, AS, AR, AT, AB, SR, ST, SB, RT, RB, TT, TB, BB |
| $\bar{4}$ | $\bar{4} \parallel Z_3$ | No | No | AA, AS, AR, AT, AB, SR, ST, SB, RT, RB, TT, TB, BB |
| $4/m$ | $4 \parallel Z_3, m \perp Z_3$ | No | No | AA, AS, AR, SR, TT, TB, BB |
| $422$ | $4 \parallel Z_3, 2 \parallel Z_1$ | Yes | No | AA, AT, SR, SB, RB, TT, BB |
| $4mm$ | $4 \parallel Z_3, m \perp Z_1$ | No | No | AA, AB, SR, ST, SB, RT, RB, TT, BB |
| $\bar{4}2m$ | $\bar{4} \parallel Z_3, 2 \parallel Z_1$ | No | No | AA, AT, SR, SB, RB, TT, BB |
| $4/mmm$ | $4 \parallel Z_3, m \perp Z_3, m \perp Z_1$ | No | No | AA, SR, TT, BB |
| $3$ | $3 \parallel Z_3$ | Yes | No | AA, AS, AR, AT, AB, SS, SR, ST, SB, RT, RB, TT, TB, BB |
| $\bar{3}$ | $\bar{3} \parallel Z_3$ | No | Yes | AA, AS, AR, SS, SR, TT, TB, BB |
| $32$ | $3 \parallel Z_3, 2 \parallel Z_1$ | Yes | No | AA, AS, AR, AT, AB, SS, SR, ST, SB, RT, RB, TT, TB, BB |
| $3m$ | $3 \parallel Z_3, m \perp Z_1$ | No | No | AA, AS, AR, AB, SS, SR, ST, SB, RT, RB, TT, TB, BB |
| $\bar{3}m$ | $\bar{3} \parallel Z_3, m \perp Z_1$ | No | Yes | AA, AS, AR, SS, SR, TT, TB, BB |
| $6$ | $6 \parallel Z_3$ | Yes | No | AA, AR, AT, AB, SR, ST, SB, RT, RB, TT, TB, BB |
| $\bar{6}$ | $\bar{6} \parallel Z_3$ | No | No | AA, AR, AB, SR, ST, SB, RT, RB, TT, TB, BB |
| $6/m$ | $6 \parallel Z_3, m \perp Z_3$ | No | Yes | AA, AR, SR, TT, TB, BB |
| $622$ | $6 \parallel Z_3, 2 \parallel Z_1$ | Yes | No | AA, AT, SR, SB, RB, TT, BB |
| $6mm$ | $6 \parallel Z_3, m \perp Z_1$ | No | No | AA, AB, SR, ST, SB, RT, RB, TT, BB |
| $\bar{6}m2$ | $\bar{6} \parallel Z_3, m \perp Z_1$ | No | No | AA, AB, SR, ST, SB, RT, RB, TT, BB |
| $6/mmm$ | $6 \parallel Z_3, m \perp Z_3, m \perp Z_1$ | No | No | AA, SR, TT, BB |
| 23 | $2 \parallel Z_1, 3 \parallel [1\,1\,1]$ | No | No | AA, AT, SR, SB, RB, TT, BB |
| $m\bar{3}$ | $m \perp Z_1, 3 \parallel [1\,1\,1]$ | No | Yes | AA, SR, TT, BB |
| 432 | $4 \parallel Z_3, 3 \parallel [1\,1\,1]$ | No | No | AA, AT, SR, SB, RB, TT, BB |
| $\bar{4}3m$ | $\bar{4} \parallel Z_3, 3 \parallel [1\,1\,1]$ | No | No | AA, AT, SR, SB, RB, TT, BB |
| $m\bar{3}m$ | $m \perp Z_1, 3 \parallel [1\,1\,1], m \perp [1\,1\,0]$ | No | Yes | AA, TT, BB |

To meet the limited length requirements of this article, we select 5 point groups out of a total of 32 for 3D lattices (Table 3–7) and analyze coupling effects from the decoupled elasticity tensor **q**. We present **q** of the remaining 27 point groups in the **Appendix**. Notably, an isotropic elastic tensor is not included in the 32 point groups of the 3D crystallography; we present a Curie point group in the Cauchy elasticity with infinite rotational symmetry representing an isotropy tensor [19] in Table 8.

Note that the nonzero components in the elasticity tensors can be zero for specific geometries; the point group is a necessary but not sufficient condition of the corresponding coupling effects [19]. In addition, because the lattices of the 32 point groups are not isotropic, the elasticity tensors and the potential coupling effects can vary with a coordinate transformation.

Table 3 shows the decoupled micropolar elasticity tensor **q** of point group 1 and the corresponding coupling effects. Because the only symmetry operation in point group 1 is identity, it is a complete anisotropy. The number of independent components in **q** is 171, the largest among the 32 point groups. The potential coupling effects include AA, AS, AR, AT, AB, SS, SR, ST, SB, RR, RT, RB, TT, TB, and BB coupling, i.e., all the 15 coupling effects in 3D lattices.

*Table 3. Decoupled micropolar elasticity tensor **q** of point group* 1

$$\begin{bmatrix}
\begin{matrix} c_{11} & c_{12} & c_{13} \\ & c_{22} & c_{23} \\ \mathbf{AA} & & c_{33} \end{matrix} &
\begin{matrix} c_{14} & c_{15} & c_{16} \\ c_{24} & c_{25} & c_{26} \\ c_{34} & c_{35} & c_{36} \end{matrix} &
\begin{matrix} c_{17} & c_{18} & c_{19} \\ c_{27} & c_{28} & c_{29} \\ c_{37} & c_{38} & c_{39} \end{matrix} &
\begin{matrix} b_{11} & b_{12} & b_{13} \\ b_{21} & b_{22} & b_{23} \\ b_{31} & b_{32} & b_{33} \end{matrix} &
\begin{matrix} b_{14} & b_{15} & b_{16} \\ b_{24} & b_{25} & b_{26} \\ b_{34} & b_{35} & b_{36} \end{matrix} &
\begin{matrix} b_{17} & b_{18} & b_{19} \\ b_{27} & b_{28} & b_{29} \\ b_{37} & b_{38} & b_{39} \end{matrix} \\
\mathbf{AS} &
\begin{matrix} c_{44} & c_{45} & c_{46} \\ & c_{55} & c_{56} \\ \mathbf{SS} & & c_{66} \end{matrix} &
\begin{matrix} c_{47} & c_{48} & c_{49} \\ c_{57} & c_{58} & c_{59} \\ c_{67} & c_{68} & c_{69} \end{matrix} &
\begin{matrix} b_{41} & b_{42} & b_{43} \\ b_{51} & b_{52} & b_{53} \\ b_{61} & b_{62} & b_{63} \end{matrix} &
\begin{matrix} b_{44} & b_{45} & b_{46} \\ b_{54} & b_{55} & b_{56} \\ b_{64} & b_{65} & b_{66} \end{matrix} &
\begin{matrix} b_{47} & b_{48} & b_{49} \\ b_{57} & b_{58} & b_{59} \\ b_{67} & b_{68} & b_{69} \end{matrix} \\
\mathbf{AR} & \mathbf{SR} &
\begin{matrix} c_{77} & c_{78} & c_{79} \\ & c_{88} & c_{89} \\ \mathbf{RR} & & c_{99} \end{matrix} &
\begin{matrix} b_{71} & b_{72} & b_{73} \\ b_{81} & b_{82} & b_{83} \\ b_{91} & b_{92} & b_{93} \end{matrix} &
\begin{matrix} b_{74} & b_{75} & b_{76} \\ b_{84} & b_{85} & b_{86} \\ b_{94} & b_{95} & b_{96} \end{matrix} &
\begin{matrix} b_{77} & b_{78} & b_{79} \\ b_{87} & b_{88} & b_{89} \\ b_{97} & b_{98} & b_{99} \end{matrix} \\
\mathbf{AT} & \mathbf{ST} & \mathbf{RT} &
\begin{matrix} d_{11} & d_{12} & d_{13} \\ & d_{22} & d_{23} \\ \mathbf{TT} & & d_{33} \end{matrix} &
\begin{matrix} d_{14} & d_{15} & d_{16} \\ d_{24} & d_{25} & d_{26} \\ d_{34} & d_{35} & d_{36} \end{matrix} &
\begin{matrix} d_{17} & d_{18} & d_{19} \\ d_{27} & d_{28} & d_{29} \\ d_{37} & d_{38} & d_{39} \end{matrix} \\
\mathbf{AB} & \mathbf{SB} & \mathbf{RB} & \mathbf{TB} &
\begin{matrix} d_{44} & d_{45} & d_{46} \\ & d_{55} & d_{56} \\ \mathbf{BB} & & d_{66} \end{matrix} &
\begin{matrix} d_{47} & d_{48} & d_{49} \\ d_{57} & d_{58} & d_{59} \\ d_{67} & d_{68} & d_{69} \end{matrix} \\
\mathbf{AB} & \mathbf{SB} & \mathbf{RB} & \mathbf{TB} & \mathbf{BB} &
\begin{matrix} d_{77} & d_{78} & d_{79} \\ & d_{88} & d_{89} \\ \mathbf{BB} & & d_{99} \end{matrix}
\end{bmatrix}$$

Table 4 shows **q** of point group $\bar{1}$ to identify the coupling effects. The symmetry operations of point group $\bar{1}$ are identity (1) and inversion ($\bar{1}$), where we omit the identity in the rest because it exists in every point group. Compared with **q** of point group 1, all $b_{ij}$ components become zero, i.e., the inversion symmetry operation ($\bar{1}$) leads to zero components for the tensor **b**, which is consistent with previous work [31]. Because all $b_{ij}$ are zero, the AT, AB, ST, SB, RT, and RB couplings no longer exist. The possible coupling effects of point group $\bar{1}$ are AA, AS, AR, SS, SR, RR, TT, TB, and BB couplings, as indicated in Table 4. The number of independent components is 90. Lakes and Benedict [32] showed that non-centrosymmetry with nonzero $b_{ij}$ leads to an AT coupling. Note that **q** in Table 4 is centrosymmetric ($b_{ij} = 0$), not having an AT coupling, which is consistent with their conclusion.

*Table 4. Decoupled micropolar elasticity tensor **q** of point group $\bar{1}$; symmetry operation - $\bar{1}$*

$$\begin{bmatrix}
\begin{matrix} c_{11} & c_{12} & c_{13} \\ & c_{22} & c_{23} \\ \mathbf{AA} & & c_{33} \end{matrix} &
\begin{matrix} c_{14} & c_{15} & c_{16} \\ c_{24} & c_{25} & c_{26} \\ c_{34} & c_{35} & c_{36} \end{matrix} &
\begin{matrix} c_{17} & c_{18} & c_{19} \\ c_{27} & c_{28} & c_{29} \\ c_{37} & c_{38} & c_{39} \end{matrix} &
\begin{matrix} 0 & 0 & 0 \\ 0 & 0 & 0 \\ 0 & 0 & 0 \end{matrix} &
\begin{matrix} 0 & 0 & 0 \\ 0 & 0 & 0 \\ 0 & 0 & 0 \end{matrix} &
\begin{matrix} 0 & 0 & 0 \\ 0 & 0 & 0 \\ 0 & 0 & 0 \end{matrix} \\
\mathbf{AS} &
\begin{matrix} c_{44} & c_{45} & c_{46} \\ & c_{55} & c_{56} \\ \mathbf{SS} & & c_{66} \end{matrix} &
\begin{matrix} c_{47} & c_{48} & c_{49} \\ c_{57} & c_{58} & c_{59} \\ c_{67} & c_{68} & c_{69} \end{matrix} &
\begin{matrix} 0 & 0 & 0 \\ 0 & 0 & 0 \\ 0 & 0 & 0 \end{matrix} &
\begin{matrix} 0 & 0 & 0 \\ 0 & 0 & 0 \\ 0 & 0 & 0 \end{matrix} &
\begin{matrix} 0 & 0 & 0 \\ 0 & 0 & 0 \\ 0 & 0 & 0 \end{matrix} \\
\mathbf{AR} & \mathbf{SR} &
\begin{matrix} c_{77} & c_{78} & c_{79} \\ & c_{88} & c_{89} \\ \mathbf{RR} & & c_{99} \end{matrix} &
\begin{matrix} 0 & 0 & 0 \\ 0 & 0 & 0 \\ 0 & 0 & 0 \end{matrix} &
\begin{matrix} 0 & 0 & 0 \\ 0 & 0 & 0 \\ 0 & 0 & 0 \end{matrix} &
\begin{matrix} 0 & 0 & 0 \\ 0 & 0 & 0 \\ 0 & 0 & 0 \end{matrix} \\
& & &
\begin{matrix} d_{11} & d_{12} & d_{13} \\ & d_{22} & d_{23} \\ \mathbf{TT} & & d_{33} \end{matrix} &
\begin{matrix} d_{14} & d_{15} & d_{16} \\ d_{24} & d_{25} & d_{26} \\ d_{34} & d_{35} & d_{36} \end{matrix} &
\begin{matrix} d_{17} & d_{18} & d_{19} \\ d_{27} & d_{28} & d_{29} \\ d_{37} & d_{38} & d_{39} \end{matrix} \\
& & & \mathbf{TB} &
\begin{matrix} d_{44} & d_{45} & d_{46} \\ & d_{55} & d_{56} \\ \mathbf{BB} & & d_{66} \end{matrix} &
\begin{matrix} d_{47} & d_{48} & d_{49} \\ d_{57} & d_{58} & d_{59} \\ d_{67} & d_{68} & d_{69} \end{matrix} \\
& & & \mathbf{TB} & \mathbf{BB} &
\begin{matrix} d_{77} & d_{78} & d_{79} \\ & d_{88} & d_{89} \\ \mathbf{BB} & & d_{99} \end{matrix}
\end{bmatrix}$$

Table 5 shows **q** and the coupling effects of point group $mm2$. Point group $mm2$ has a mirror symmetry operation perpendicular to the $Z_1$ and $Z_2$ axes. There are 50 independent components in **q** and 9 possible coupling effects — AA, AB, SR, ST, SB, RT, RB, TT, and BB couplings. All components of AA and TT couplings in **q** are non-zeros. However, there are partially zero components in the AB, SR, ST, SB, RT, RB, and BB couplings.

*Table 5. Decoupled micropolar elasticity tensor **q** of point group $mm2$; symmetry operation - $m \perp Z_1$, $m \perp Z_2$*

$$\begin{bmatrix}
\begin{array}{ccc} c_{11} & c_{12} & c_{13} \\ & c_{22} & c_{23} \\ \textbf{AA} & & c_{33} \end{array} &
\begin{array}{ccc} 0 & 0 & 0 \\ 0 & 0 & 0 \\ 0 & 0 & 0 \end{array} &
\begin{array}{ccc} 0 & 0 & 0 \\ 0 & 0 & 0 \\ 0 & 0 & 0 \end{array} &
\begin{array}{ccc} 0 & 0 & 0 \\ 0 & 0 & 0 \\ 0 & 0 & 0 \end{array} &
\begin{array}{ccc} 0 & 0 & b_{16} \\ 0 & 0 & b_{26} \\ 0 & 0 & b_{36} \end{array} &
\begin{array}{ccc} 0 & 0 & b_{19} \\ 0 & 0 & b_{29} \\ 0 & 0 & b_{39} \end{array} \\
& \begin{array}{ccc} c_{44} & 0 & 0 \\ & c_{55} & 0 \\ & & c_{66} \end{array} &
\begin{array}{ccc} c_{47} & 0 & 0 \\ 0 & c_{58} & 0 \\ c_{67} & 0 & c_{69} \end{array} &
\begin{array}{ccc} 0 & 0 & 0 \\ 0 & 0 & 0 \\ b_{61} & b_{62} & b_{63} \end{array} &
\begin{array}{ccc} 0 & b_{45} & 0 \\ b_{54} & 0 & 0 \\ 0 & 0 & 0 \end{array} &
\begin{array}{ccc} 0 & b_{48} & 0 \\ b_{57} & 0 & 0 \\ 0 & 0 & 0 \end{array} \\
& & \begin{array}{c} \textbf{SR} \end{array} \quad \begin{array}{ccc} c_{77} & 0 & 0 \\ & c_{88} & 0 \\ & & c_{99} \end{array} &
\begin{array}{ccc} 0 & 0 & 0 \\ 0 & 0 & 0 \\ b_{91} & b_{92} & b_{93} \end{array} &
\begin{array}{ccc} 0 & b_{75} & 0 \\ b_{84} & 0 & 0 \\ 0 & 0 & 0 \end{array} &
\begin{array}{ccc} 0 & b_{78} & 0 \\ b_{87} & 0 & 0 \\ 0 & 0 & 0 \end{array} \\
& \textbf{ST} & \textbf{RT} & \begin{array}{ccc} d_{11} & d_{12} & d_{13} \\ & d_{22} & d_{23} \\ \textbf{TT} & & d_{33} \end{array} &
\begin{array}{ccc} 0 & 0 & 0 \\ 0 & 0 & 0 \\ 0 & 0 & 0 \end{array} &
\begin{array}{ccc} 0 & 0 & 0 \\ 0 & 0 & 0 \\ 0 & 0 & 0 \end{array} \\
\textbf{AB} & \textbf{SB} & \textbf{RB} & & \begin{array}{ccc} d_{44} & 0 & 0 \\ & d_{55} & 0 \\ & & d_{66} \end{array} &
\begin{array}{ccc} d_{47} & 0 & 0 \\ 0 & d_{58} & 0 \\ 0 & 0 & d_{69} \end{array} \\
\textbf{AB} & \textbf{SB} & \textbf{RB} & & \textbf{BB} & \begin{array}{ccc} d_{77} & 0 & 0 \\ & d_{88} & 0 \\ & & d_{99} \end{array}
\end{bmatrix}$$

Table 6 shows **q** of point group $422$ and the corresponding couplings. The symmetry operations of point group $422$ are a four-fold rotational symmetry parallel to the $Z_3$ axis and a two-fold rotational symmetry parallel to the $Z_1$ axis. There are 29 independent components in **q** and seven possible coupling effects — AA, AT, SR, SB, RB, TT, and BB couplings. The AA, AT, and TT couplings have all nonzero components in **q**, while partially nonzero components affect SR, SB, RB, and BB couplings.

*Table 6. Decoupled micropolar elasticity tensor **q** of point group $422$; Symmetry operation: $4 \parallel Z_3$, $2 \parallel Z_1$*

$$\begin{bmatrix}
\begin{array}{ccc} c_{11} & c_{12} & c_{13} \\ & c_{11} & c_{13} \\ \text{AA} & & c_{33} \end{array} &
\begin{array}{ccc} 0 & 0 & 0 \\ 0 & 0 & 0 \\ 0 & 0 & 0 \end{array} &
\begin{array}{ccc} 0 & 0 & 0 \\ 0 & 0 & 0 \\ 0 & 0 & 0 \end{array} &
\begin{array}{ccc} b_{11} & b_{12} & b_{13} \\ b_{12} & b_{11} & b_{13} \\ b_{31} & b_{31} & b_{33} \end{array} &
\begin{array}{ccc} 0 & 0 & 0 \\ 0 & 0 & 0 \\ 0 & 0 & 0 \end{array} &
\begin{array}{ccc} 0 & 0 & 0 \\ 0 & 0 & 0 \\ 0 & 0 & 0 \end{array} \\
& \begin{array}{ccc} c_{44} & 0 & 0 \\ & c_{44} & 0 \\ \square & & c_{66} \end{array} &
\begin{array}{ccc} c_{47} & 0 & 0 \\ 0 & -c_{47} & 0 \\ 0 & 0 & 0 \end{array} &
\begin{array}{ccc} 0 & 0 & 0 \\ 0 & 0 & 0 \\ 0 & 0 & 0 \end{array} &
\begin{array}{ccc} b_{44} & 0 & 0 \\ 0 & b_{44} & 0 \\ 0 & 0 & b_{66} \end{array} &
\begin{array}{ccc} b_{47} & 0 & 0 \\ 0 & b_{47} & 0 \\ 0 & 0 & b_{66} \end{array} \\
& & \begin{array}{ccc} c_{77} & 0 & 0 \\ & c_{77} & 0 \\ \square & & c_{99} \end{array} &
\begin{array}{ccc} 0 & 0 & 0 \\ 0 & 0 & 0 \\ 0 & 0 & 0 \end{array} &
\begin{array}{ccc} b_{74} & 0 & 0 \\ 0 & -b_{74} & 0 \\ 0 & 0 & b_{96} \end{array} &
\begin{array}{ccc} b_{77} & 0 & 0 \\ 0 & -b_{77} & 0 \\ 0 & 0 & -b_{96} \end{array} \\
\text{AT} & & & \begin{array}{ccc} d_{11} & d_{12} & d_{13} \\ & d_{11} & d_{13} \\ \text{TT} & & d_{33} \end{array} &
\begin{array}{ccc} 0 & 0 & 0 \\ 0 & 0 & 0 \\ 0 & 0 & 0 \end{array} &
\begin{array}{ccc} 0 & 0 & 0 \\ 0 & 0 & 0 \\ 0 & 0 & 0 \end{array} \\
& \text{SB} & \text{RB} & & \begin{array}{ccc} d_{44} & 0 & 0 \\ & d_{44} & 0 \\ \square & & d_{66} \end{array} &
\begin{array}{ccc} d_{47} & 0 & 0 \\ 0 & d_{47} & 0 \\ 0 & 0 & d_{69} \end{array} \\
& \text{SB} & \text{RB} & & \text{BB} & \begin{array}{ccc} d_{77} & 0 & 0 \\ & d_{77} & 0 \\ \square & & d_{66} \end{array}
\end{bmatrix}$$

Table 7 shows **q** and the coupling effects of point group 432. The symmetry operations of point group 432 are a four-fold rotational symmetry parallel to the $Z_3$ axis and a three-fold rotational symmetry parallel to [1 1 1]. There are 12 independent components in **q** and 7 possible coupling effects — AA, AT, SR, SB, RB, TT, and BB couplings. Similar to point group 422 in Table 6, point group 432 has all nonzero components in **q** for the AA, AT, and TT couplings, while it has partially nonzero components for the SR, SB, RB, and BB couplings.

*Table 7. Decoupled micropolar elasticity tensor **q** of point group 432; Symmetry operation: 4 ∥ $Z_3$, 3 ∥ [1 1 1]*

$$\begin{bmatrix}
\begin{array}{ccc} c_{11} & c_{12} & c_{12} \\ & c_{11} & c_{12} \\ \text{AA} & & c_{11} \end{array} &
\begin{array}{ccc} 0 & 0 & 0 \\ 0 & 0 & 0 \\ 0 & 0 & 0 \end{array} &
\begin{array}{ccc} 0 & 0 & 0 \\ 0 & 0 & 0 \\ 0 & 0 & 0 \end{array} &
\begin{array}{ccc} b_{11} & b_{12} & b_{12} \\ b_{12} & b_{11} & b_{12} \\ b_{12} & b_{12} & b_{11} \end{array} &
\begin{array}{ccc} 0 & 0 & 0 \\ 0 & 0 & 0 \\ 0 & 0 & 0 \end{array} &
\begin{array}{ccc} 0 & 0 & 0 \\ 0 & 0 & 0 \\ 0 & 0 & 0 \end{array} \\
& \begin{array}{ccc} c_{44} & 0 & 0 \\ & c_{44} & 0 \\ \square & & c_{44} \end{array} &
\begin{array}{ccc} c_{47} & 0 & 0 \\ 0 & c_{47} & 0 \\ 0 & 0 & c_{47} \end{array} &
\begin{array}{ccc} 0 & 0 & 0 \\ 0 & 0 & 0 \\ 0 & 0 & 0 \end{array} &
\begin{array}{ccc} b_{44} & 0 & 0 \\ 0 & b_{44} & 0 \\ 0 & 0 & b_{44} \end{array} &
\begin{array}{ccc} b_{44} & 0 & 0 \\ 0 & b_{44} & 0 \\ 0 & 0 & b_{44} \end{array} \\
& & \begin{array}{ccc} c_{77} & 0 & 0 \\ \text{SR} & c_{77} & 0 \\ \square & & c_{77} \end{array} &
\begin{array}{ccc} 0 & 0 & 0 \\ 0 & 0 & 0 \\ 0 & 0 & 0 \end{array} &
\begin{array}{ccc} b_{74} & 0 & 0 \\ 0 & -b_{74} & 0 \\ 0 & 0 & b_{74} \end{array} &
\begin{array}{ccc} -b_{74} & 0 & 0 \\ 0 & b_{74} & 0 \\ 0 & 0 & -b_{74} \end{array} \\
\text{AT} & & & \begin{array}{ccc} d_{11} & d_{12} & d_{12} \\ & d_{11} & d_{12} \\ \text{TT} & & d_{11} \end{array} &
\begin{array}{ccc} 0 & 0 & 0 \\ 0 & 0 & 0 \\ 0 & 0 & 0 \end{array} &
\begin{array}{ccc} 0 & 0 & 0 \\ 0 & 0 & 0 \\ 0 & 0 & 0 \end{array} \\
& \text{SB} & \text{RB} & & \begin{array}{ccc} d_{44} & 0 & 0 \\ & d_{44} & 0 \\ \square & & d_{44} \end{array} &
\begin{array}{ccc} d_{47} & 0 & 0 \\ 0 & d_{47} & 0 \\ 0 & 0 & d_{47} \end{array} \\
& \text{SB} & \text{RB} & & \text{BB} & \begin{array}{ccc} d_{44} & 0 & 0 \\ & d_{44} & 0 \\ \square & & d_{44} \end{array}
\end{bmatrix}$$

The point group theory classification can be applied to any periodic 3D lattice. For example, the rigid, compliant, auxetic, and chiral lattices in [33] belong to point groups $4/m$, $m3m$, $m3m$, and 432, respectively. However, the previous works did not investigate the mechanical coupling effects of these lattices. Several 3D lattice materials have been designed to demonstrate an AT coupling [9,13,24]; both belong to point group 432, supported by Table 7 with nonzero $b_{ij}$. Note that some previous geometry [34,35] demonstrating an AT coupling is not periodic lattices where one cannot describe the behavior with the point group theory.

Notably, none of the elasticity tensors in the 32 point groups is isotropic, which is consistent with Cauchy elasticity [19]. The isotropic micropolar elasticity tensor is well known [36,37], which can be transformed into the decoupled form, and the result is shown in Table 8. The elasticity tensor has six independent variables and three possible coupling effects — AA, TT, and BB couplings. Because this is the simplest form of the micropolar elasticity tensor, it is fair to say that AA, TT, and BB couplings are the inherent nature of the micropolar continuum, just like Poisson's effect in the Cauchy elasticity.

Table 8. Decoupled isotropic micropolar elasticity tensor $q$, where $c_{44} = c_{11} - c_{12}/2$ and $d_{47} = d_{11} - d_{12} - d_{44}$

$$\begin{bmatrix}
c_{11} & c_{12} & c_{12} & 0 & 0 & 0 & 0 & 0 & 0 & 0 & 0 & 0 & 0 & 0 & 0 & 0 & 0 & 0 \\
 & c_{11} & c_{12} & 0 & 0 & 0 & 0 & 0 & 0 & 0 & 0 & 0 & 0 & 0 & 0 & 0 & 0 & 0 \\
\textbf{AA} & & c_{11} & 0 & 0 & 0 & 0 & 0 & 0 & 0 & 0 & 0 & 0 & 0 & 0 & 0 & 0 & 0 \\
 & & & c_{44} & 0 & 0 & 0 & 0 & 0 & 0 & 0 & 0 & 0 & 0 & 0 & 0 & 0 & 0 \\
 & & & & c_{44} & 0 & 0 & 0 & 0 & 0 & 0 & 0 & 0 & 0 & 0 & 0 & 0 & 0 \\
 & & & & & c_{44} & 0 & 0 & 0 & 0 & 0 & 0 & 0 & 0 & 0 & 0 & 0 & 0 \\
 & & & & & & c_{77} & 0 & 0 & 0 & 0 & 0 & 0 & 0 & 0 & 0 & 0 & 0 \\
 & & & & & & & c_{77} & 0 & 0 & 0 & 0 & 0 & 0 & 0 & 0 & 0 & 0 \\
 & & & & & & & & c_{77} & 0 & 0 & 0 & 0 & 0 & 0 & 0 & 0 & 0 \\
 & & & & & & & & & d_{11} & d_{12} & d_{12} & 0 & 0 & 0 & 0 & 0 & 0 \\
 & & & & & & & & & & d_{11} & d_{12} & 0 & 0 & 0 & 0 & 0 & 0 \\
 & & & & & & & & \textbf{TT} & & & d_{11} & 0 & 0 & 0 & 0 & 0 & 0 \\
 & & & & & & & & & & & & d_{44} & 0 & 0 & d_{47} & 0 & 0 \\
 & & & & & & & & & & & & & d_{44} & 0 & 0 & d_{47} & 0 \\
 & & & & & & & & & & & & & & d_{44} & 0 & 0 & d_{47} \\
 & & & & & & & & & & & & & & & d_{44} & 0 & 0 \\
 & & & & & & & & & & & & & \textbf{BB} & & & d_{44} & 0 \\
 & & & & & & & & & & & & & & & & & d_{44}
\end{bmatrix}$$

## V. Verification with FE simulations

Figure 4 shows an AS coupling of a curved cubic lattice belonging to point group $\bar{1}$ where the decoupled micropolar elasticity tensor is found in Table 4. We apply a compressive loading of $\varepsilon_{33} = -0.1$ on the rigid top plate while fixing the rigid plate on the bottom, ending up with an AS coupling by shear deformations in the $Z_1$ and $Z_2$ directions. $c_{34}$, $c_{35}$, and $c_{36}$ in Table 4 are the components affecting the AS coupling by an axial loading in the $Z_3$ direction. Similarly, the curved cubic lattice can generate an AS coupling by an axial loading in the $Z_1$ direction with $c_{14}$, $c_{15}$, and $c_{16}$ in Table 4. An axial loading in $Z_2$ can also produce an AS coupling by $c_{24}$, $c_{25}$, and $c_{26}$ in Table 4.

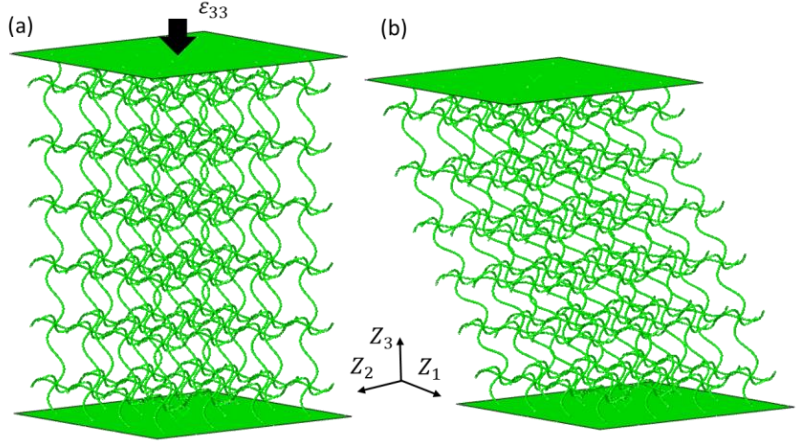

*Figure 4. Axial–shear (AS) coupling of $\bar{1}$ lattice: (a) initial geometry and (b) deformed shape.*

Figure 5 shows a BT coupling of the cubic lattice of point group $\bar{1}$; applying bending $\kappa_{31}$ on the top and bottom plates produce a twisting deformation $\kappa_{33}$ affected by $d_{35}$ in Table 4. Note that $d_{34}$, $d_{35}$, and $d_{36}$ in Table 4 are the components to provide a BT coupling of $\kappa_{33}$. In addition, the BT and TB couplings are identical due to the mutual coupling. Notably, the $\bar{1}$ lattice has both AS and BT couplings, meaning that the single structure generates a multimodal coupling. The multimodal coupling is significant in the design of sensors and actuators; without a complicated combination of machine components, one can generate both shear and twisting deformations with selective input loading modes.

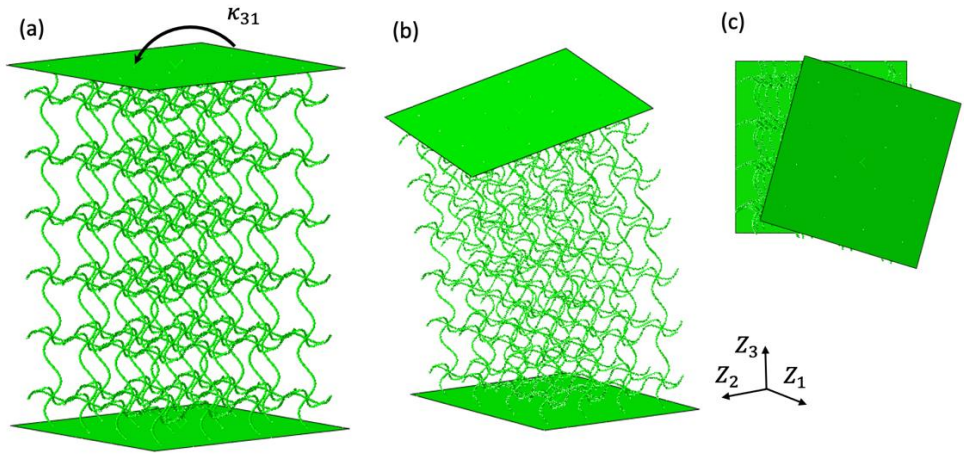

*Figure 5. Bending–twisting (BT) coupling of $\bar{1}$ lattice: (a) initial geometry, (b) deformed shape, and (c) top view of deformed shape.*

Figure 6 shows an AB coupling of point group $mm2$; applying a compressive loading of $\varepsilon_{33} = -0.05$ while fixing the bottom ligaments can generate a bending $\kappa_{21}$ affected by $b_{39}$ in Table 4. Note that $b_{39}$ in Table 4 affects a bending of $\kappa_{12}$ by an input loading $\varepsilon_{33}$. Similarly, $b_{16}$ and $b_{19}$ affect $\kappa_{12}$ and $\kappa_{21}$ by an input loading $\varepsilon_{11}$. An input axial loading $\varepsilon_{22}$ causes bending deformations $\kappa_{12}$ and $\kappa_{21}$ by $b_{26}$ and $b_{29}$, respectively. Replacing two straight struts with a single curvature wavy shape in a 2D square lattice could produce an AB coupling [11]. We can extend this principle to the 3D lattices by replacing four straight struts with a single curved one, as shown in Figure 6. Note that this deformation is not by buckling and there is no shear deformation by the axial loading input $\varepsilon_{33}$. By replacing the straight beams with curved ones, one can generate an AB coupling with different point groups, e.g., the point groups $mm2$, $m$, 2, 1, 4, $\bar{4}$, and $4mm$.

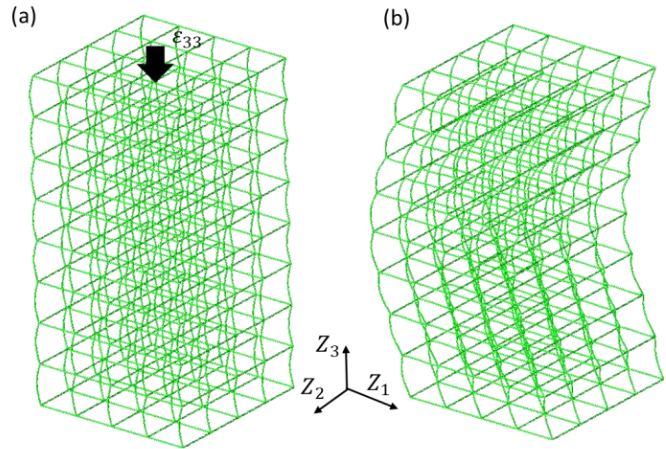

*Figure 6. Axial–bending (AB) coupling of $mm2$ lattice: (a) initial geometry and (b) deformed shape.*

Figure 7 shows a TA coupling of a curved 3D lattice belonging to point group 422; applying a compressive loading of $\varepsilon_{33} = -0.05$ on the top plate while fixing the bottom plate produces a twisting deformation $\kappa_{33}$ affected by $b_{33}$ in Table 6. Note that the non-zero $b_{31}$ and $b_{32}$ cause $\kappa_{11}$ and $\kappa_{22}$ by an axial input loading of $\varepsilon_{33}$. This AT coupling was previously also identified by several groups under chiral design [9,24]. However, their designs belong to point group 432 without a chiral effect.

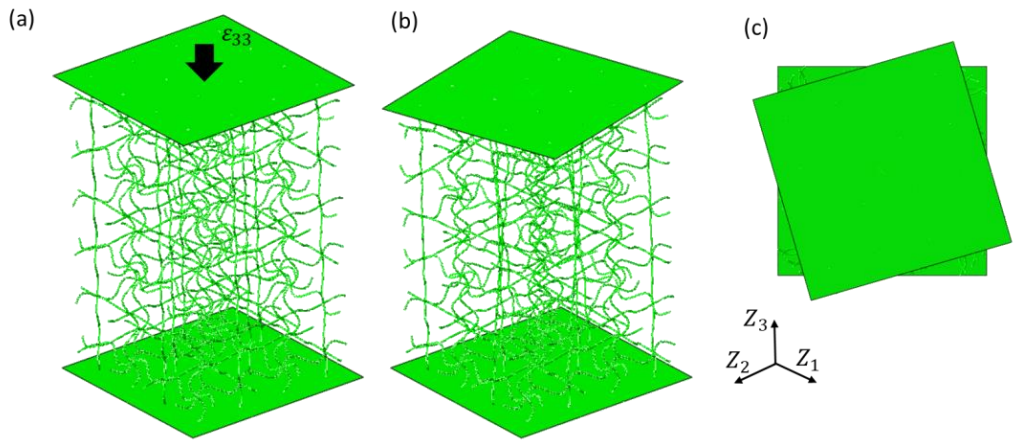

*Figure 7. Axial–twisting (AT) coupling of 422 lattice: (a) initial geometry, (b) deformed shape, (c) top view of deformed shape.*

Figure 8 demonstrates an SB coupling of a curved 3D lattice belonging to point group 422; applying a pure shear loading of $\varepsilon_{13} = -0.05$ causes a bending deformation $\kappa_{13}$ affected by $b_{44}$ in Table 6. Similarly, shear loadings $\varepsilon_{23}$ and $\varepsilon_{12}$ can produce bending couplings $\kappa_{23}$ and $\kappa_{12}$, respectively, affected by $b_{44}$ and $b_{66}$.

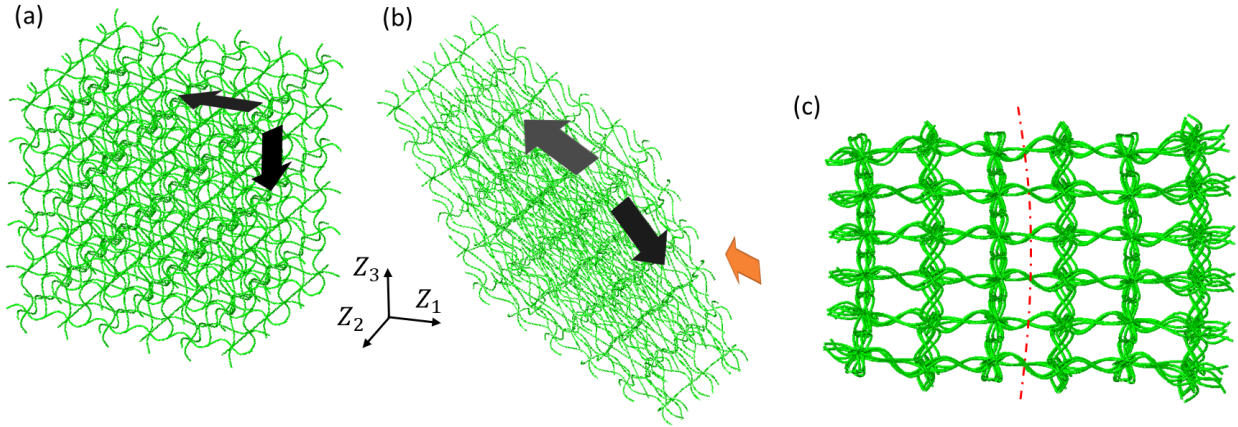

*Figure 8. Shear–bending (SB) coupling of a $422$ lattice: (a) initial geometry, (b) deformed shape, and (c) directional view of deformed shape observed toward the orange arrow in (b). The black arrows indicate the applied shear loading.*

Similar to point group $\bar{1}$, point group 422 has multimodal coupling effects. Previously, a lattice belonging to point group 432 demonstrated only AT coupling [9,24] even though the lattice structure has a multiple coupling opportunity. According to our analysis, a lattice of point group 432 has AA, AT, SR, SB, RB, TT, and BB couplings, as shown in Table 2, which warrants future investigation.

## VI. Discussion

With the development of the advanced manufacturing technique, artificial structural materials with complex 3D geometry are becoming significant due to the potential to overcome the limited physical properties of conventional materials. The previous ad hoc designs of mechanical coupling of architected materials without a comprehensive symmetry analysis [9,24] do not provide robust design guidelines for mechanical couplings. Therefore, we construct the elasticity tensor covering comprehensive mechanical couplings of 3D lattices. The decoupled micropolar elasticity tensor **q** provides a total of 15 couplings. The comprehensive '**q**'s of 32 point groups offer potential mechanical couplings with geometric symmetry of 3D lattice materials, significantly contributing to the design of architected materials.

Compared with 2D lattices' eight couplings of four modes – axial, shear, bending, and rotation [23], 3D lattices in this study offer 15 couplings with an added twisting mode and in-plane/out-of-plane couplings of the existing four modes. The decoupled micropolar elasticity tensor **q** in 3D lattices exceeds the 2D counterpart in size and number; a matrix form of $18 \times 18$ with 32 point groups in 3D over a matrix form of $6 \times 6$ with 10 point groups in 2D [23]. Micropolar elasticity indeed provides a superior number of coupling effects; the anisotropic Cauchy elasticity only provides three couplings – AA, AS, and SS [14]. Alternatively, CLT provides several couplings with a broken symmetry of layups – AS, AB, AT, SB, ST, BB, and BT couplings [16]. However, CLT still does not cover pure in-plane couplings, lacking in description of the whole coupling effects of 3D structures. Moreover, the rotation mode in the micropolar elasticity generates unique couplings over CLT – **SR**, **RR**, **RT**, and **RB** couplings.

The classification of geometry with chirality and non-centrosymmetry cannot fully describe the whole couplings, providing limited couplings, e.g., an AT coupling in 3D lattices [9,24]. The integration of the point group theory with the micropolar elasticity in this study can open a new era for a vast horizon of mechanical couplings. Our study opens a new branch in the exploration of mechanical couplings with geometric symmetry in 3D metamaterials design, greatly expanding the design space of lattice materials with significant applications in sensors, actuators, mechanical logic gates, acoustic cloaking, elastic waveguides, as well as many other areas.

## VII. Conclusion

In summary, we developed a generalized methodology to provide comprehensive mechanical couplings in 3D lattices using the decoupled micropolar elasticity tensor obtained by decomposing shear and rotational deformations. Following Neumann's principle, we constructed the decoupled elasticity tensors classified into 32 individual point groups, providing potential mechanical couplings with lattice geometries. The decoupled micropolar elasticity tensor offers a total of 15 independent coupling effects of 3D lattices. The seven couplings — AR, SR, RR, RT, RB, TT, and BB couplings — are uniquely identified by the micropolar elasticity. FE simulations of mechanical couplings of selective 3D lattices provide a good demonstration of our methodology. The decoupled micropolar elasticity tensors for each point group provide solid theoretical guidelines on the mechanical couplings of 3D structural materials with potential applications in various areas, including active metamaterials, sensors, actuators, elastic waveguides, and acoustics.

## VIII. Acknowledgments

The authors acknowledge support received from the Shanghai NSF (Award # 17ZR1414700) and the Research Incentive Program of Recruited Non-Chinese Foreign Faculty by Shanghai Jiao Tong University.